\documentclass[graybox]{svmult}
\usepackage{mathptmx}       
\usepackage{helvet}         
\usepackage{courier}        
\usepackage{type1cm}        
\usepackage{subfigure}                            

\usepackage{makeidx}         
\usepackage{graphicx}        
\usepackage{multicol}        
\usepackage[bottom]{footmisc}
\makeindex             
\begin{document}

\title{When the path is never shortest: a reality check on shortest path biocomputation}
\titlerunning{When the path is never shortest}

\author{Richard Mayne}
\institute{R. Mayne \at Unconventional Computing Laboratory, University of the West of England, Bristol, United Kingdom \\ \email{Richard.Mayne@uwe.ac.uk}}
%

%
\maketitle

\abstract{Shortest path problems are a touchstone for evaluating the computing performance and functional range of novel computing substrates. Much has been published in recent years regarding the use of biocomputers to solve minimal path problems such as route optimisation and labyrinth navigation, but their outputs are typically difficult to reproduce and somewhat abstract in nature, suggesting that both experimental design and analysis in the field require standardising. This chapter details laboratory experimental data which probe the path finding process in two single-celled protistic model organisms, \emph{Physarum polycephalum} and \emph{Paramecium caudatum}, comprising a shortest path problem and labyrinth navigation, respectively. The results presented illustrate several of the key difficulties that are encountered in categorising biological behaviours in the language of computing, including biological variability, non-halting operations and adverse reactions to experimental stimuli. It is concluded that neither organism examined are able to efficiently or reproducibly solve shortest path problems in the specific experimental conditions that were tested. Data presented are contextualised with biological theory and design principles for maximising the usefulness of experimental biocomputer prototypes. }

\section{Introduction}

This chapter addresses the use of biocomputer prototypes for addressing various minimal path problems (MPPs), which include physical solving of graph theoretical tasks such as shortest path problems (SPPs; synonymous with calculation of the Steiner minimum spanning tree of a set of vertices), the Travelling Salesperson Problem (TSP) and labyrinth navigation, by living systems.  Measurement of an organism's ability to solve such puzzles, especially mazes, is not new: rodent navigation through geometrically-confined spaces was a staple of psychological research during the previous century and various forms of maze puzzle remain a diverting brain-teaser for children. It has not been until the comparatively recent advent of digital computing, however, that we have begun to question the practical applications for experimental biocomputing prototypes that are able to address graph theory problems.

Of particular note is the explosion in research on the navigational abilities of the macroscopic amoeba-like organism \emph{Physarum polycephalum} during the past decade which have experimentally demonstrated that this single-celled organism is capable of solving the TSP \cite{Zhu2013}, navigating through various labyrinths and geometric puzzles on the first pass \cite{Nakagaki2000,Reid2012} and calculating minimum spanning tree of a series of vertices \cite{Adamatzky2007}, all via adaptation of its somatic morphology as a result of foraging behaviours (which will be expanded upon in the following section). Other notable examples of graphical biocomputation from recent years include, but are not limited to:

\begin{enumerate}
\item TSP solving through induced genetic transformation within live bacterial cells, wherein edges are conceptualised as segments of DNA linking gene nodes. Despite the variety of means by which computation can be achieved in transgenic bacteria, output is usually interpreted optically, e.g. through a change in colony colour (expression of coloured/fluorescent proteins), or expression of antibiotic resistance genes \cite{Baumgardner2009,Esau2014}.
\item Ant swarm migration along optimised single pathways --- deduced by pathfinder ants according to an edge weight of attractant and repellent gradients --- towards new nesting sites \cite{Pratt2002}. These dynamics may be put to more tangible computing applications such as addressing the Towers of Hanoi problem \cite{Reid2011}.
\item Maze navigation via the shortest path by cultured epithelial tumour cells, apparently guided by self-generated chemical gradients in a manner suggested to underlie cancer cell invasion \cite{Scherber2012}.
\item Navigation through complex virtual reality labyrinths by rats undergoing simultaneous neural measurement \cite{Harvey2009}.
\end{enumerate}

Research in this area of biocomputation is justified for the following reasons. Firstly, as we approach the limitations of conventional digital hardware (i.e. the finite nature of the miniaturisation barrier presented by silicon-based computing substrates and associated problem of waste energy thermalisation), we are led to question the value of novel substrates, techniques and applications for computing technologies. Secondly, as alluded to in point 3 of the above enumerated list, interpretation of natural behaviours as expressions of computing aid our understanding of the biosciences and by extension, our ability to experimentally manipulate them. Finally, development of bio-inspired algorithms for use on conventional computing architectures is a richly diverse and varied area of research with virtually limitless applications; examples of successful algorithms relevant to this chapter include ant colony systems for optimised calculation of a range of problems including the TSP, labyrinth navigation and foraging route optimisation \cite{Dorigo1997,Vittori2006,Vela-Perez2013,Ramsch2012} and multi-agent \emph{P. polycephalum}-inspired models for solving SPPs and hence planning transport networks \cite{Jones2015}. 

This apparently glowing appraisal of MPP biocomputation (also called `bioevaluation'), however, somewhat misrepresents the abilities of biological organisms for approaching problems that we are accustomed to tackling via the use of conventional computers, i.e. machines operating according to principles of the Turing model. As was eloquently argued by Stepney \cite{Stepney2008}, biological substrates can only be said to compute in a distinctly non-Turing fashion, i.e. they are non-designed entities that are, to all intents and purposes, non-halting, nonsymbolic,  stochastic systems. We are, furthermore, currently far from having elucidated the biological processes (intracellular, intercellular and extracellular signalling events) that constitute biocomputer input/output operations and interactions therein that we are choosing to call a form of computation. As such, the majority of experimental biocomputer prototypes will suffer from poor reproducibility, be slow\footnote{Biological time is many orders of magnitude slower than electrical time, implying that biocomputers capitalise on their comparative parallelism in order to beat the efficiency of conventional substrates.}, costly to operate and require large amounts of operator time investment to set up, program and monitor for output. All of these factors are distinctly far-removed from our usual conception of computation. How, then, can live substrates be said to `compute' the solution to MPPs given their aforementioned detriments? 

The purpose of this chapter is to examine two case studies documenting research which exemplifies the theoretical and experimental limitations of utilising biocomputing substrates for calculating MPPs and enforce a `reality check' on MPP biocomputation in so doing. In plainer terms, the aim of this chapter is to delineate the differences in the way in which biological substrates can be said to `compute' the solution to MPPs, in comparison to the electronic substrate (algorithmic) equivalent. The conclusions drawn highlight the comparative strengths and weaknesses of biocomputing substrates and suggest experimental considerations for designing MPP-oriented biocomputers.

\section{Case study 1: The \emph{Physarum} problem}

\subsection{Background}
\label{sec-phyBackground}
\begin{figure}
\centering
\includegraphics[width=0.85\textwidth]{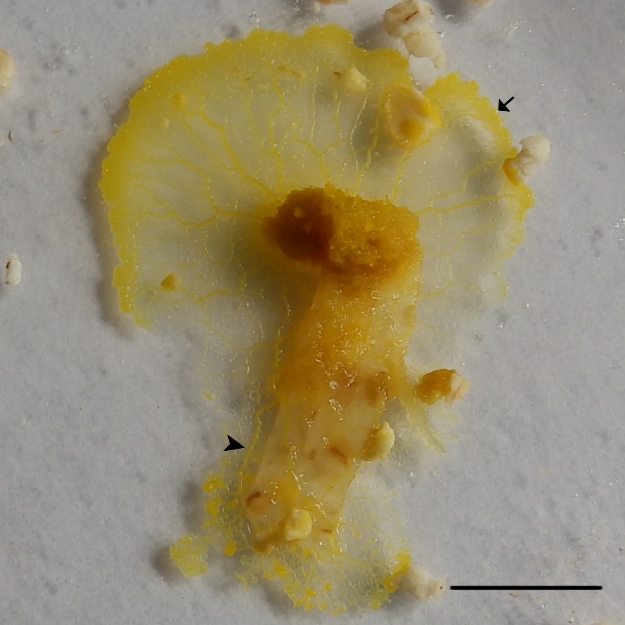}
\caption{Photograph of a \emph{P. polycephalum} plasmodium cultivated on 2\% non-nutrient agar gel, engulfing a few oat flakes. The organism is composed of caudal tubular regions (arrowhead) and a fan-shaped advancing anterior margin (arrow). Scale bar 10mm.}
\label{fig-physarum}
\end{figure}

The \emph{P. polycephalum} plasmodium (vegetative life cycle form) (Fig. \ref{fig-physarum}) is a remarkable and fascinating protistic creature that is, at the time of writing, one of the most intensively researched-upon biological computing substrates. Comprising a macroscopic amoeba-like cell possessing millions of nuclei encapsulated within a single cell membrane, the plasmodial (or `acellular') slime moulds are archetypal model organisms for excitable, motile cells as well as a go-to organism for educators wishing to demonstrate simple culture techniques with non-pathogenic organisms. 

It was discovered in 2000 that \emph{P. polycephalum} could navigate through a maze puzzle on the first pass \cite{Nakagaki2000}: this precipitated a biocomputing revolution\footnote{A Google Scholar (\url{http://scholar.google.com}) query with the search terms `Physarum' and `computing' return approximately 2,750 hits between the years 2000--2017 (search date September 2017).} that saw the development of slime mould sensors, computer interfaces and circuitry, to name but a few examples of what are informally known as `Physarum machines'. Whilst it is inappropriate to expand further on the range of biocomputing applications that have been found for \emph{P. polycephalum}, we refer the reader to Refs. \cite{Adamatzky2010,Adamatzky2016,Mayne2016}, and Jones \& Safonow's chapter in this volume, for a comprehensive overview of slime mould computing. 

As maze navigation is a MPP, this novel experiment quickly led researchers to question what other problems of graph theory that could be applied to slime mould. In 2007, Adamatzky \cite{Adamatzky2007} demonstrated that slime mould may solve SPPs, guided by nutrient gradients, thus demonstrating a clear advantage over previous reaction-diffusion computing substrates which cannot address this class of problem without engineered collisions. Slime mould mechanisms for adapting its inbuilt foraging behaviour to solving permutations of SPPs have been exploited in various biocomputer prototypes, including:

\begin{itemize}
\item Solving U-shaped trap problems \cite{Reid2012}.
\item Colour sensing \cite{Adamatzky2013}.
\item Various logic gates \cite{Mayne2015}.
\item Constructing proximity graphs and beta skeletons \cite{Adamatzky2009b}.
\item Designing transport networks \cite{Tero2010}.
\item Constructing the convex and concave hulls about spatially-distributed nutrient sources \cite{Jones2015b}.
\end{itemize}     

The abilities of slime mould are represented with different degrees of `enthusiasm' by the researchers who them: some maintain cautiously that slime mould biocomputation of MPPs are `approximations' and hence that we are assigning conventional computing terminology on the organism for ease of comparison \cite{Adamatzky2010}, whereas at the other end of the spectrum, some authors claim that the organism is capable of `intelligent' behaviour \cite{Nakagaki2000}. 

In this section, I report that although slime mould SPP bioevaluation is both fascinating and worthy of further research, the manner in which the organism is able to represent the shape of a dataset is incompatible with our familiar, algorithmic understanding of the concept of `solving a SPP'.

The experiments outlined in this section rely on observing the migration of slime mould through a down-scaled two-dimensional representation of a human living space, guided by chemoattractant gradients and zones of photorepulsion. The specific application of these experiments was to `bio-evaluate' the layout of this domicile with regard to how `efficiently' the space is subdivided, but their design is essentially the same as all of the previously mentioned Physarum machines created to address MPPs, i.e. analysis of the organism's migration between spatially-distributed nutrient sources. Success of the organism with regards to SPP navigation was interpreted in terms of the following criteria:

\begin{enumerate}
\item[a.] Ability of the organism to navigate between a finite number of attractant sources (vertices) in an order that represents a shortest path solution. Results were compared output from the same problem when addressed by a conventional computer (using Dijkstra's algorithm), i.e. if the shortest path between a set of four vertices in a virtual two-dimensional space, [A,B,C,D], is solved (via calculation of edge weights) by a computer as [D] $\rightarrow$ [B] $\rightarrow$ [A] $\rightarrow$ [C], slime mould in corresponding experiments whose conditions mimic those in the simulation will be judged to have correctly calculated the SPP if it navigates between vertices in the order [D,B,A,C]. This is opposed to simply comparing edge lengths between laboratory experiments (physically measuring the slime mould's length) and computer simulations.
\item[b.] Reproducibility, i.e. the ability laboratory experimental slime mould to consistently navigate the same route between distributed vertices.
\end{enumerate}

This methodology was chosen to best represent the differences between the way in which SPPs are `bioevaluated', rather than algorithmically `computed'. Total edge length travelled by slime mould in each experiment was also measured (for comparison but not as a primary determinant of path navigation `success'), as were general observations on culture morphology and behaviour. 

The rationale of this experiment was that the organism was expected to plot a route through the living space: we expect, informed by previous work on the topic, that the slime mould will attempt to link the discrete nutrient sources by forming tubular strands of protoplasm between them in a shape approximating the shortest path between them, whilst avoiding illuminated areas (hence why the shortest path solution outcome is not measured here as purely the total edge length). 

The application of this Physarum machine was therefore to grant insight into a natural route around the space, e.g. if the organism were to visit the kitchen first, we could reason that this is not the most efficient use of space as the kitchen may not be the most frequently-visited room or indeed the room one is most likely to visit on entering the building. Although this application is somewhat removed from computer science, it represents an active area of unconventional computing, i.e. `bioinspiration', which looks to analyse natural behaviour and apply it to creative problems; its inclusion here is purely to exemplify how the organism calculates its path.

\subsection{Methods}

A sample of a \emph{P. polycephalum} plasmodium colonising an oat flake (a preferred slime mould nutrient source) was placed onto a section of 2\% non-nutrient agarose gel. In this experiment, the agar section (hereafter `wet layer') represents a geometric environment (or, graph) wherein vertices in the forthcoming SPP were represented by oat flakes. The wet layer was situated within a 900~mm square plastic Petri dish made of clear polystyrene.

\begin{figure}
\centering
\includegraphics[width=0.85\textwidth]{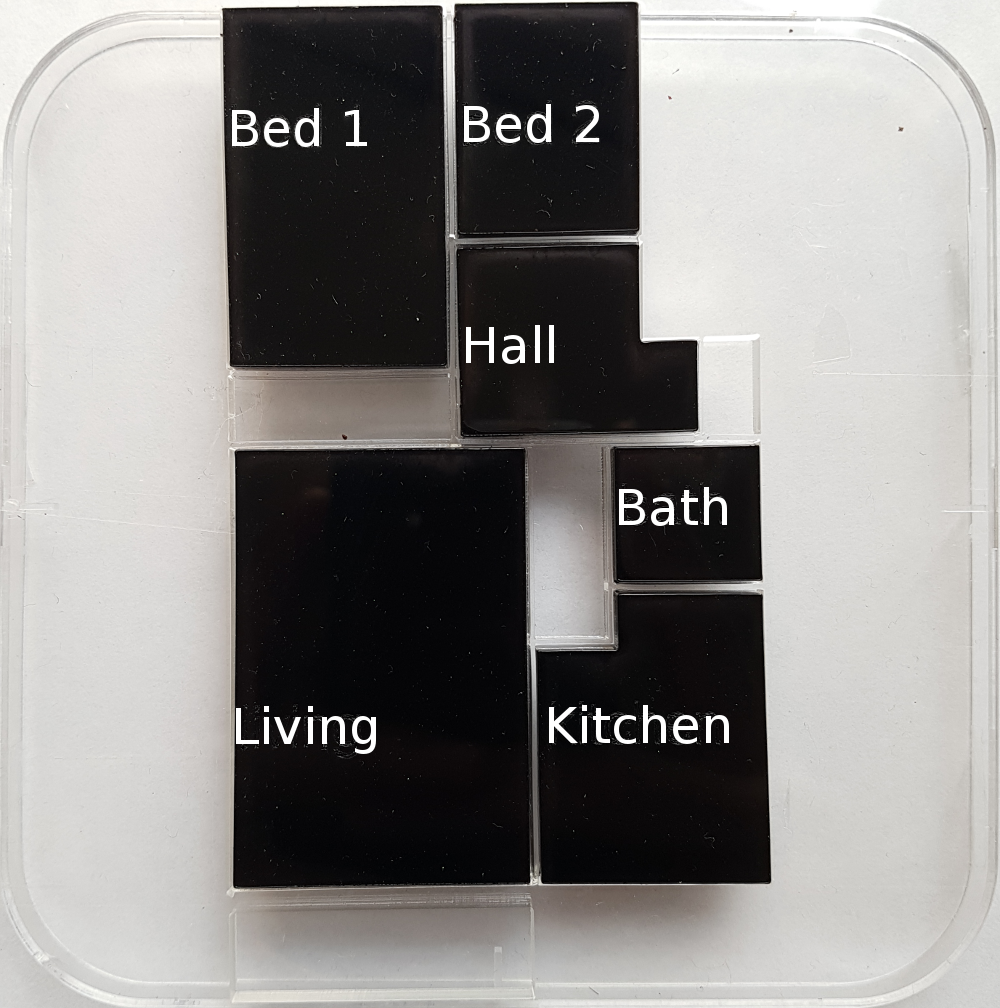}
\caption{Photograph of laser cut acrylic representation of a two bedroom, single floor living space, referred to in text as the `dry layer'. All named rooms are cut from opaque acrylic and the spaces between them (walls and hallways) are clear acrylic.}
\label{fig-domicile}
\end{figure}

A two-dimensional spatial representation of a living space was constructed as follows. An architectural draft of a two bedroom flat was etched onto two pieces of acrylic, one clear and one opaque. Both pieces of acrylic were then laser cut around each `room' (two bedrooms, living space, kitchen, bathroom, entrance hallway and two connecting hallways), leaving two `backing layers' with holes in and two sets of eight cut-out pieces. The opaque cut outs were then slotted into the clear background, with the exception of the hallways, resulting in the draft comprising opaque sections (rooms) with clear sections separating them (walls and hallways) respectively (hereafter, the acrylic portion is known as the `dry layer') (Fig. \ref{fig-domicile}).  

The dry layer was affixed to the bottom of the Petri dish housing the wet layer and the whole environment was placed overlying an electroluminescent plate producing 196~Lux, in order to represent the two-dimensional living space in a format the slime mould could interpret:  opaque `room' sections of the dry layer cast shadows onto the wet layer, whereas the clear spaces were illuminated, thus creating the organism's preferred dark zones and repellent illuminated zones. The agar gel in the wet layer was cut to the outline of the draft to constrain the organism's movements to within the `building' and uncolonised oat flakes (discrete chemoattractant sources) were arranged to sit in the centre of each room, with the initial colonisation point being placed in the space representing the entrance hallway. The experiment was shielded from external light sources and the organism was left to propagate around its environment, with photographs being taken every 4 hours for 48 hours. The experiment was repeated in triplicate.

\subsection{Results}

\begin{figure}
\centering
\subfigure[]{\includegraphics[width=0.85\textwidth]{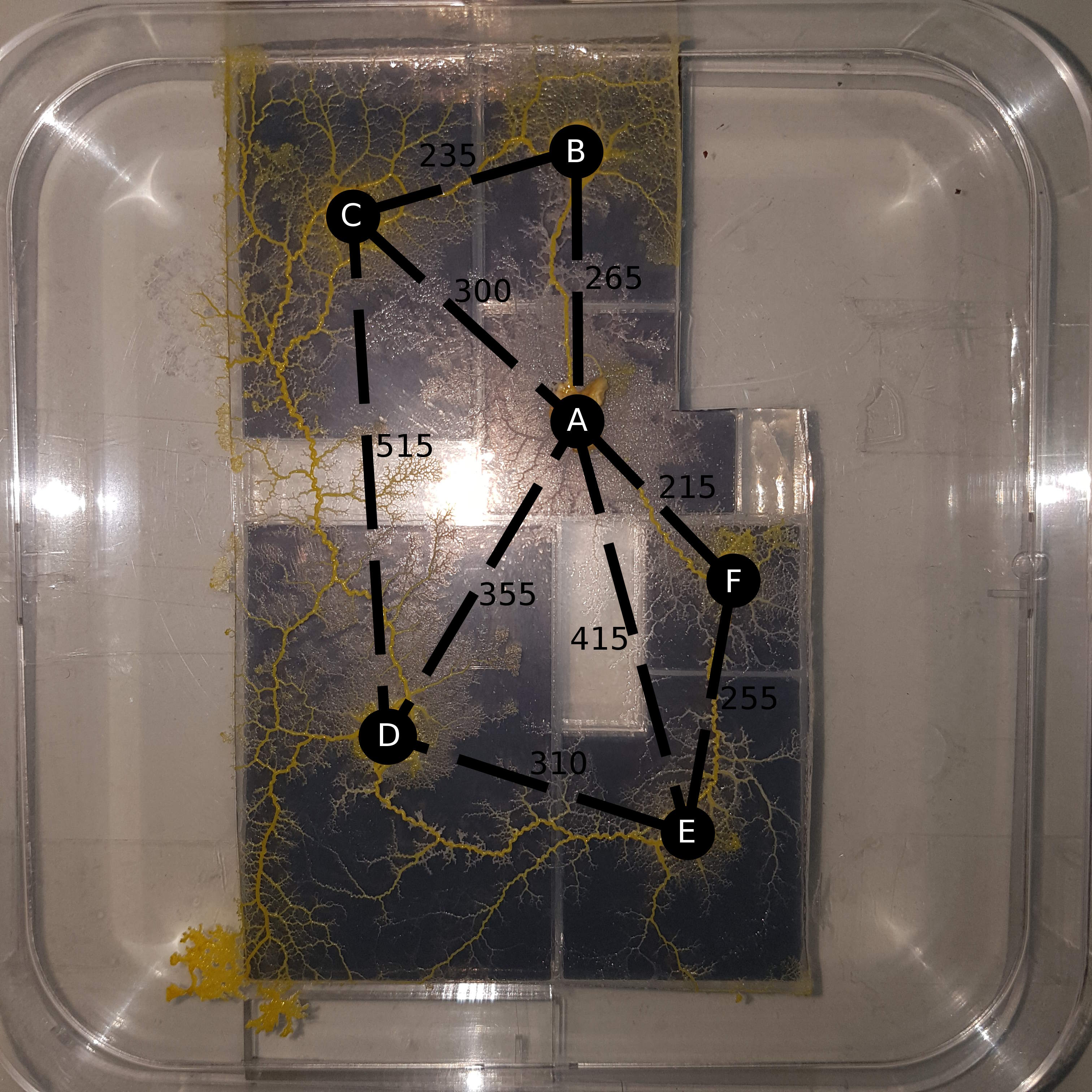}}
\subfigure[]{\includegraphics[width=0.42\textwidth]{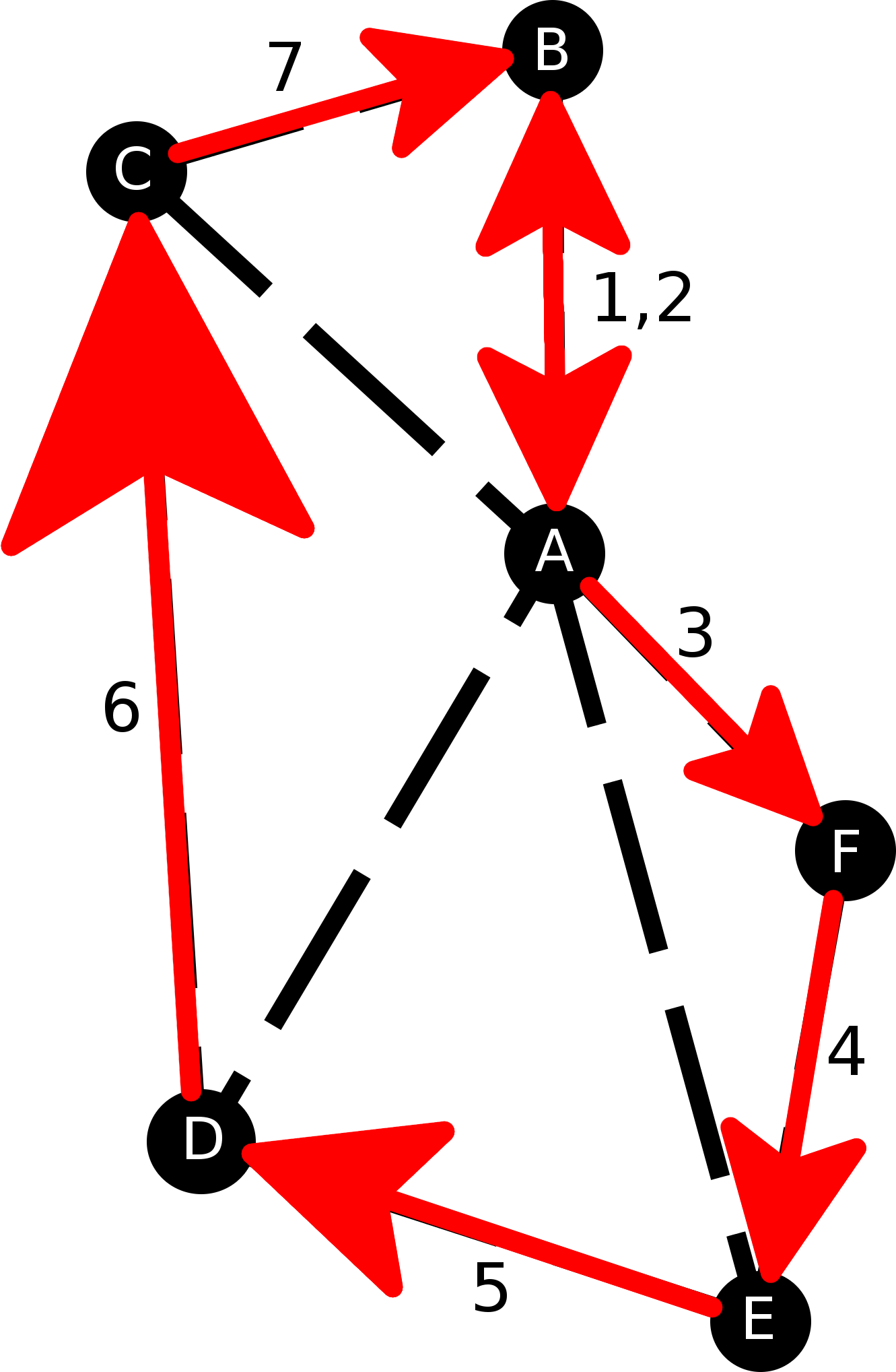}}
\subfigure[]{\includegraphics[width=0.42\textwidth]{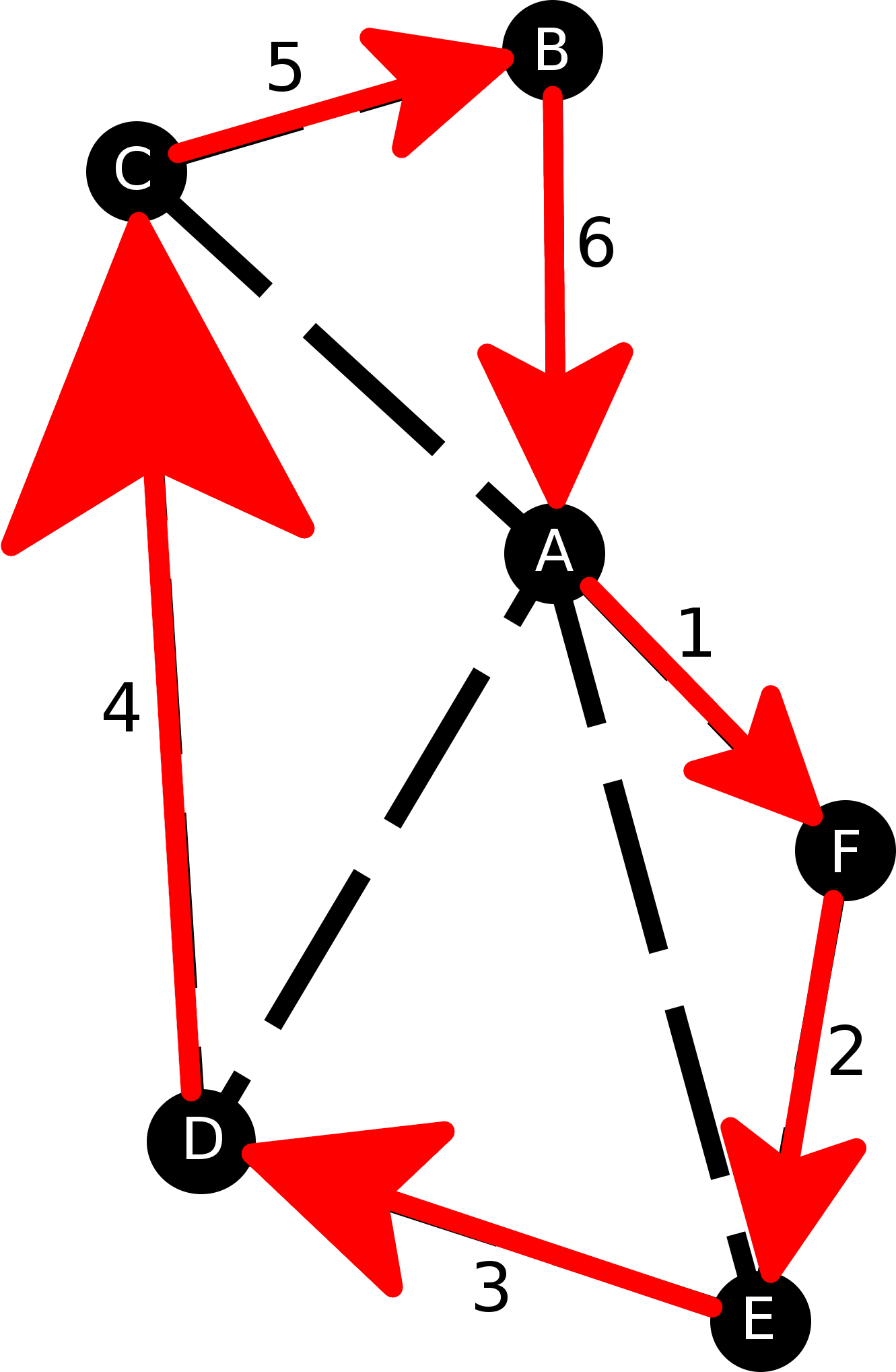}}
\caption{Slime mould addresses a SPP in a simple graph guided by attractants and repellents. (a) Photograph showing experiment (details in text) after 48 hours, overlaid with paths radiating from inoculation point, [A]. Path lengths in millimetres, rounded to nearest 5. (b) Path taken by slime mould, as indicated by arrows and numbering. (c) Shortest path, as calculated by Dijkstra's algorithm.}
\label{fig-paths}
\end{figure}

A completed representative experiment, photographed after 48~hours had elapsed, is shown in Fig. \ref{fig-paths}a. Path lengths radiating from the inoculation point [A] are overlaid; paths not shown are (in mm): [BD] 600, [BE] 675, [BF] 445, [CE] 685, [CF] 515, [DF] 375. The route taken by the plasmodium is shown in Fig. \ref{fig-paths}b and follows [A]$\rightarrow$[B]$\rightarrow$[A]$\rightarrow$[F]$\rightarrow$[E]$\rightarrow$[D]$\rightarrow$[C]$\rightarrow$[B] to complete a circuit linking all of the oat flakes. The total length of this specific route is 2060~mm, although the physical length of the organism greatly exceeds this (see below). For comparison, the minimal length solution to this problem, as calculated by Dijkstra's algorithm, is shown in \ref{fig-paths}c and follows [A]$\rightarrow$[F]$\rightarrow$[E]$\rightarrow$[D]$\rightarrow$[C]$\rightarrow$[B], with a total path length of 1795~mm.

Morphologically, the plasmodium shown in Fig. \ref{fig-paths}a contains a number of redundant links: this is best exemplified in the organism's [CD] link, which vaguely resembles the figure `8' due to two bifurcations which concatenate at the nodes. As was mentioned in the previous paragraph, the organism's travel distance between nodes exceeds the algorithmically-generated value, due to the organism not propagating in straight lines and the existence of multiple accessory branches in its protoplasmic network. The organism's deviation from straight lines tended to increase proportionally with the inter-node distance and larger `rooms' tended to have more accessory branches within them. The organism also tended to avoid crossing illuminated zones but did cross 5 wall spaces and one hallway.
 
In all experiments, the slime mould assumed a ring around the points in the manner shown in Fig. \ref{fig-paths}a (i.e. a concave hull around the vertices), but did not always take the same route: the other two routes taken were [AFEFABCDEF] and [ABCDEFA] (data not shown).

It is clear from the examplar data that the organisms did not calculate the shortest path, as defined by the criteria delineated in section \ref{sec-phyBackground}: the navigation sequence between vertices was not identical to the sequence calculated to be the shortest by Dijkstra's algorithm (although one did take an optimal route, i.e. the inverse of the algorithmically calculated route) and different routes were taken by each organism in each repeat.  

\subsection{Discussion}
\subsubsection{Straight lines, redundancy and accessory branches}
\label{sect-lines}
The experiment shown in Fig. \ref{fig-paths}a highlights why the total length of the organism's protoplasmic tube network (i.e. edge length) was not used as a primary determinant the experimental outcome: slime moulds do not travel in perfectly straight lines. In some cases, such as in the edge [AF], the fit is good but far from perfect. Conversely, [DC] edge measures about 534~mm (measuring the thickest tube only and no redundant paths), approximately 4\%  greater than the true shortest route between these paths. In comparison to conventional algorithmic approaches to solving SPPs, this amount of divergence is sufficient to negate the hypothesis that the organism could be said to be computing the `absolute' shortest path, especially as the error accumulates with each edge. Whilst it is beyond the remit of this investigation to debate at length on why slime moulds no not travel in perfectly straight lines, we may assume, parenthetically, that there is no distinct evolutionary advantage to doing so and that factors such as organisation of intracellular motile machinery and reception of diffused chemical signals involve stochastic elements (a common feature of biological processes that makes live substrates particularly challenging to \cite{Saiz2006} model). 

Another important factor influencing the organisms' total network length was the existence of the edge weighting factors other than node spacing. Although every attempt was made to control the independent variables in the experiments presented, a multitude of `background' factors that are extremely difficult to control are likely to have contributed to the variation in the organisms' routes observed across all experiments. Exemplar influencing factors such as the organisms' health and nutritional status, fluctuations in temperature, presence of microbes and distribution of moisture throughout the wet layer. It is likely that such factors also played an important role in determining the overall distribution of the organisms' protoplasm in particular areas of the wet layer, i.e. redundant links and accessory branches, which are thought to be constructed in areas of high nutrient availability and along gradients of attraction (chemical or otherwise), respectively \cite{Mayne2016advances}. As such, descriptions of slime mould addressing MPPs cannot be favourably interpreted in purely algorithmic terms, so our biocomputing vocabulary must be altered accordingly.

\subsubsection{Ordering of vertex visits}
The path the organism took in the experiment shown in Fig. \ref{fig-paths} involved doubling back, increasing its total path length by 2 edges (equating to over 10\% in terms of physical path length). This highlights how live biocomputers are always in a state of flux and require constant observation in order to assess the state of the computation; observing the output of the plasmodium at the 48 hour in Fig. \ref{fig-paths}a mark gives no indication as to how the organism's network was constructed. The nature of biological substrates is, insofar as we have ascribed an exogenous purpose to its foraging behaviour, non-halting; this could be considered both a benefit (it allows for assessment of the state of computation and overall system dynamics at any point) and a detriment (user input is required to determine when the operation has finished). 

Whilst the organism did navigate via an optimal route in one of the repeats, albeit not in the order calculated by Dijkstra's algorithm, the reproducibility of results from this small sample was poor by both biological and computing standards.


To address the question of why the slime mould in the above example took the longer path with the addition of the [ABA] diversion, one may be tempted to assume that the organism lacks the necessary `intelligence' to adequately distinugish between the benefits and detriments of the potential edges [AB] and [AF] and so opted for one at random, then found that the conditions at [B] were less favourable than at [A], so doubled back to explore other paths. Whilst this is certainly possible, it is apparent that the data are insufficient to properly measure the effect of the aforementioned non-visible weighting factors regarding environment favourability, organism status and the (likely nonlinear) relationship between these and the physical distances separating vertices. The complex interplay between attraction and repulsion is best illustrated here by the example of the [DC] path in Fig. \ref{fig-paths}a: the organism traverses the repellent `hallway' zone, presumably as the benefits of migrating across the gap outweigh the energy costs of circumventing it.

It is essentially impossible to control all of the variables in experiments such as those described here, hence variation in biocomputer must be anticipated and accounted for in experimental designs. As variation is the basis by which all organisms were able to evolve, an intuitively-designed biocomputer will capitalise on variation, despite this being anathema to the traditional concept of computing. 


\subsubsection{Shortest path approximations are only constructed in nutrient-limited environments}
\begin{figure}
\centering
\subfigure[]{\includegraphics[width=0.75\textwidth]{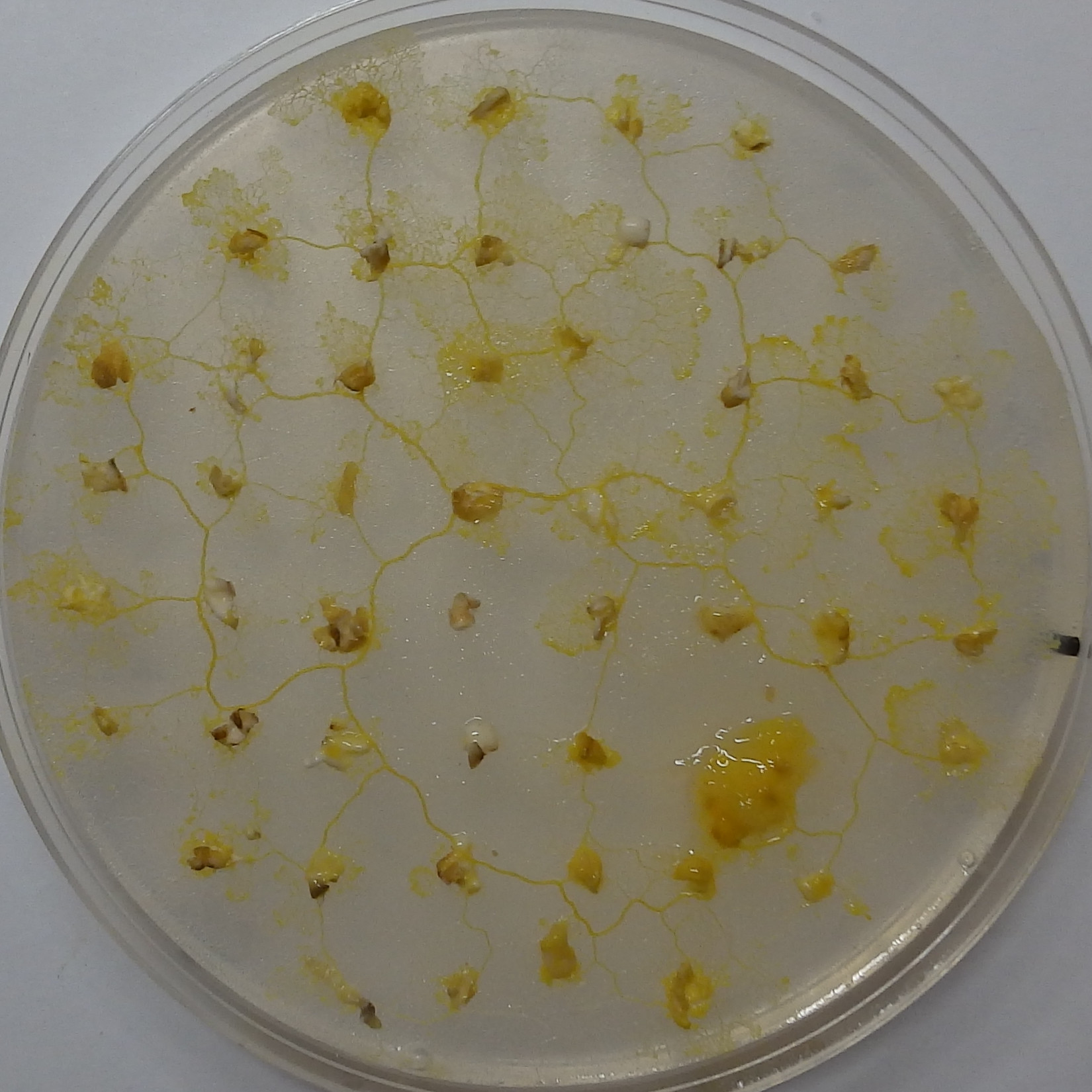}}
\subfigure[]{\includegraphics[width=0.75\textwidth]{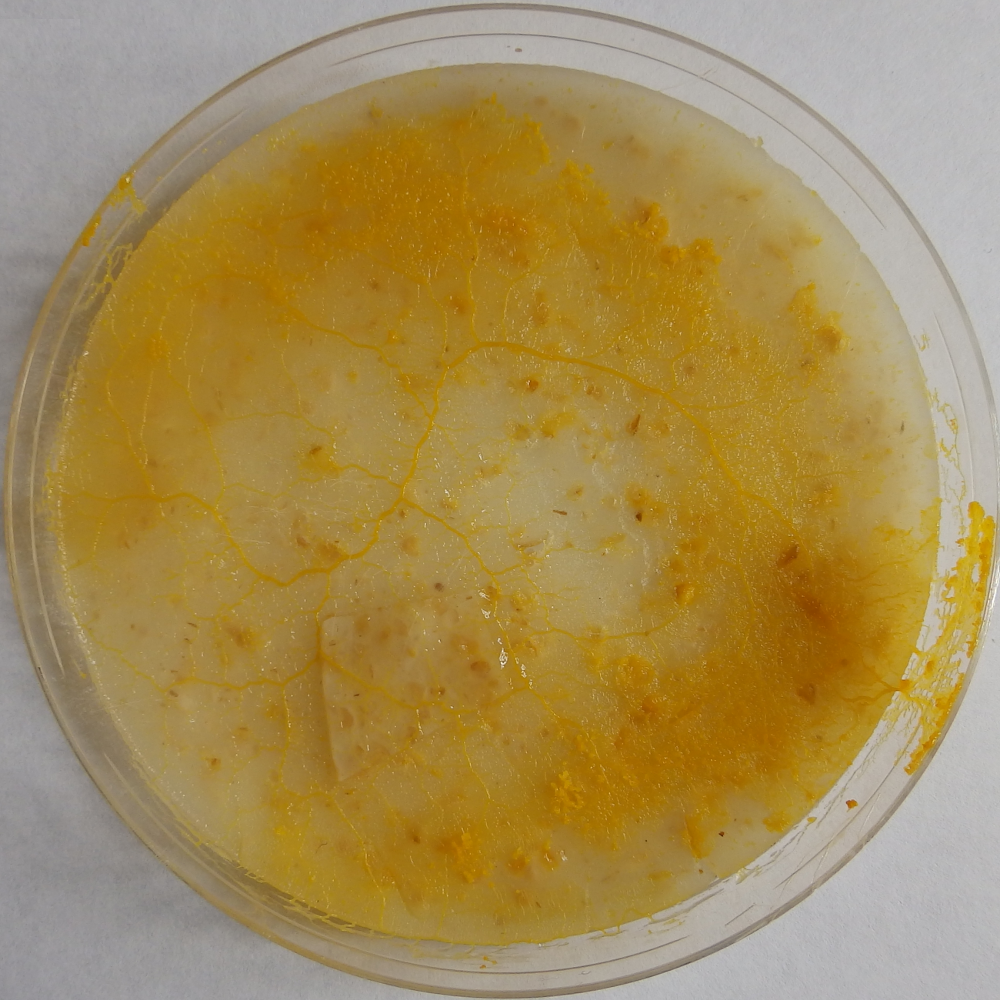}}
\caption{Photographs of \emph{P. polycephalum} propagating in nutrient rich environments, 48~h post-inoculation. (a) An excessive amount of discrete food sources (oat flakes) are provided. (b) Growing on enriched (oatmeal) agarose substrate. Adapted from \cite{Mayne2016}.}
\label{fig-physarum2}
\end{figure}

The experiment described above represents a nutrient-sparse environment for slime mould: can the organism construct similar graphs in nutrient-rich environments?  In Fig. \ref{fig-physarum2}a a \emph{P. polycephalum} plasmodium inoculated onto a lattice delineated by oat flakes is shown. The figure shows that \emph{P. polycephalum} forms a more interconnected graph in nutrient-rich environments, which is perhaps more akin to a Gabriel graph than a Steiner minimum spanning tree (although 2 points near the centre have not been linked, for an unknown reason). It was demonstrated in 2009 \cite{Adamatzky2009b} that \emph{P. polycephalum} may approximate any of the proximity graphs in the Toussaint hierarchy, dependent on relative edge weighting. For comparison, Fig. \ref{fig-physarum2}b shows slime mould growth on nutrient enriched (i.e. a uniform attractant field) agar; the entire plate is morphologically more similar to the amorphous advancing anterior margins usually observed in nutrient-limited substrates. This highlights that slime mould biocomputation of SPPs only occurs within a specific set of conditions relating to nutrient availability, meaning that reproducibility of SPPs is dependent on a fairly narrow window of initial conditions relating to the organism's nutritional status and the spacing of nutrient sources. 

\subsection{Summary}

During the recent advent of slime mould computing research, much hype was generated in the media around the use of slime mould for addressing shortest path problems, particularly with respect to route planning. The organism's malleability and easily-interpretable output led to rich and varied works, including a particularly whimsical paper in which it was suggested that slime mould should play a role in planning interplanetary missions \cite{Adamatzky2014}. In spite of this, I have demonstrated here that to label slime mould foraging behaviours in nutrient-limited environments as calculation of a SPP is somewhat inaccurate without applying a certain amount of abstraction to the manner in which the term `shortest path' is interpreted. This is not to devalue slime mould research; clearly the organism is undertaking some immensely complex massively-parallel operations, research upon which is most assuredly important. What I am suggesting, however, is that directly comparing this to conventional path finding algorithms is at best unhelpful.

\section{Case study 2: Banging your \emph{Paramecium} against a brick wall}
\subsection{Background}
\begin{figure}
\centering
\includegraphics[width=0.85\textwidth]{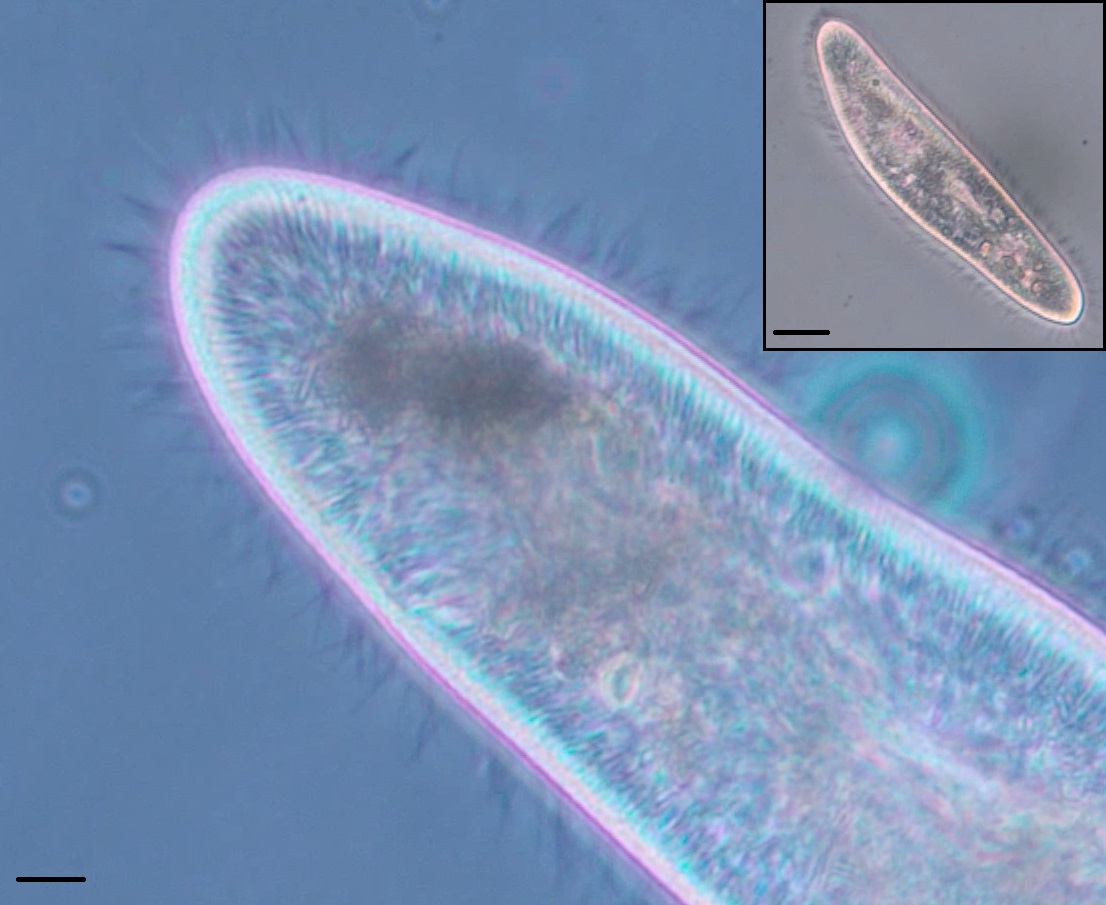}
\caption{Photomicrographs of \emph{P. caudatum}, phase contrast optics. (Main) Anterior tip of the organism, where hair-like cilia may be easily seen coating the cell's membrane. Scale bar 10~$\mu$m. (Inset) Lower magnification image showing the whole cell. Scale bar 25~$\mu$m.}
\label{fig-para}
\end{figure}

\emph{Paramecium caudatum} is a single celled protistic freshwater microorganism covered in thousands of minute hair-like appendages called `cilia' (Fig. \ref{fig-para}). Cilia beat rhythmically in order to generate fluid currents in adjacent media, thus generating motive force and enhancing feeding on dispersed particulates. Cilia-based motility in \emph{P. caudatum} therefore represents a novel mechanism for addressing MPPs in aquatic environments.

Whilst historical literature has indicated the use of basic puzzles to assess chemotaxis and theormotaxis in \emph{P. caudatum} (usually, a T-shaped puzzle where the organism is given a binary choice to navigate directly ahead or around a 90$^o$ bend) \cite{VanHouten1975}, very little attention has been paid to the organism's ability to address problems of graph theory, despite their behaviour in confined environments (microfluidic circuitry, capillary tubes) being reasonably well characterised \cite{Jana2015} \footnote{It is essential in this context to mention ingenious work of Reidel-Kruse \emph{et al.} \cite{Riedel-Kruse2011} who developed multiple `games' in tiny enclosed environments wherein paramecium behaviour was influenced by user input, including `PAC-mecium' and `ciliaball'.}.

In this section I will demonstrate how \emph{P. caudatum} is particularly ill-adapted for addressing MPPs in geometrically-constrained labyrinth puzzles and by extension illustrate some of the practical limitations of designing MPP-solving biocomputers. Single \emph{P. caudatum} cells were placed in small labyrinth puzzles in the presence of a chemoattractant gradient at the exit, according to the principle that the organism congregates in regions of highest nutrient density \cite{VanHouten1978}. Successful navigation towards the puzzle's exit within a specific timeframe (10~minutes) was judged to be evidence in support of the hypothesis that \emph{P. caudatum} are able to solve this variety of MPP.

\subsection{Methods}
\emph{P. caudatum} were cultivated in an in-house modification of Chalkley's medium which is enriched with 10~g of desiccated alfalfa and 20 grains of wheat per litre, at room temperature. Cultures were exposed to a day/night cycle but were kept out of exposed sunlight. Organisms were harvested in logarithmic growth phase by gentle centrifugation at 400$\times$G before being transferred to fresh culture media. Cells used in experiments were transferred via a micropipette to the testing environment. 

\begin{figure}
\centering
\includegraphics[width=0.85\textwidth]{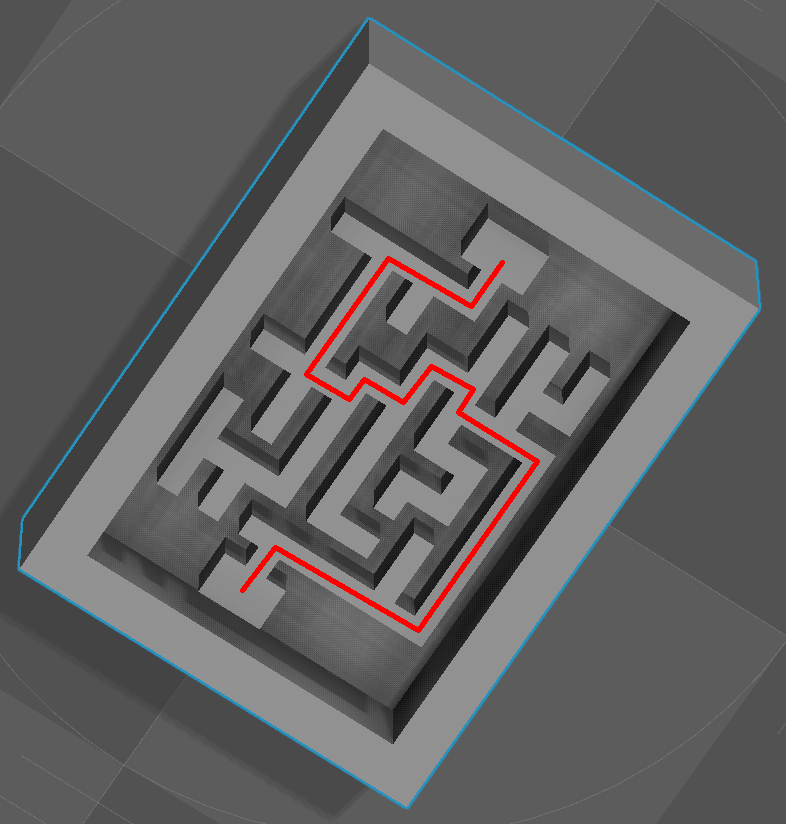}
\caption{Projection of labyrinth puzzle mould used in \emph{P. caudatum} navigation experiments. The route between the entrance/exit reservoirs is shown in red.}
\label{fig-mazeSTL}
\end{figure}

The testing environments, these being labyrinth puzzles designed to accommodate \emph{P. caudatum}, were fabricated as follows. Labyrinths were generated in openSCAD using an open-source Python script \cite{openSCAD} which rendered graphic files in STL format (Fig. \ref{fig-mazeSTL}). The size of the completed labyrinth was approximately 10$\times$7~mm; the dimensions of the labyrinth's walls were chosen to  accommodate approximately 8--10 cell widths (750~$\mu$m$^2$), which was reasoned to be ample room to allow a single \emph{P. caudatum} cell to manoeuvre and reduce the likelihood of collisions with the environment's walls. Maze designs were modified to have two distinct reservoirs at the entrance and exits to the labyrinths before being inverted using Solidwoks 2017 (Dassault Syst\`{e}mes, France). Moulds were then printed in PLA using an Objet260 FDM 3D printer (Stratasys, USA) at a resolution of 50~$\mu$m. The moulds were then cleaned in isopropyl alcohol, rinsed three times in deionised water and air dried. Labyrinths were cast in clear polydimethylsiloxane (PDMS) (Sylgard 184, Dow Corning, USA) by pouring elastomer solution onto the mould, removing any air in a vacuum chamber and finally polymerising in an oven at 40$^o$C for 48 hours.


Completed testing environments were stuck to large glass microscope coverslips (depth 0.11~mm) and a small piece of a solid chemoattractant (desiccated alfalfa) was placed in one of the labyrinth's reservoirs. The maze was then filled with tap water that had been resting for 48 hours from the end containing the solid chemoattractant source, taking care not to dislodge it from its reservoir. This method was chosen in order to generate a gradient of chemoattractants along the maze. The environment was then allowed to rest for 15 minutes in a sealed Petri dish, in order to allow bubbles to disappear but prevent fluid evaporation. Individual \emph{P. caudatum} cells were then transferred to the unoccupied reservoir using a micropipette. Observations were made using a stereomicroscope and video footage was collected using a Brunel Eyecam (Brunel Microscopy, UK). Each experiment was run for 10 minutes, after which the water began to evaporate to a noticeable degree. Labyrinths were not sealed in order to not expose the organisms to alterations in fluid pressure or dissolved oxygen content. 

The timescale for the experiment was judged to be sufficiently long for the organism to navigate the puzzle (based on typical \emph{P. caudatum} movement speed reaching in excess 1~mm per second), whilst disallowing the eventuality of the organism arriving at the exit via a random walk. The experiment was repeated 10 times.

\begin{figure}
\centering
\subfigure[]{\includegraphics[width=0.32\textwidth]{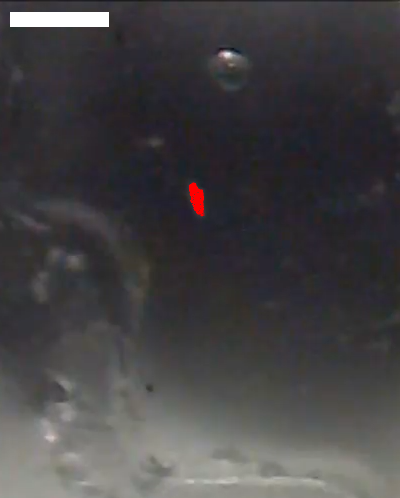}}
\subfigure[]{\includegraphics[width=0.32\textwidth]{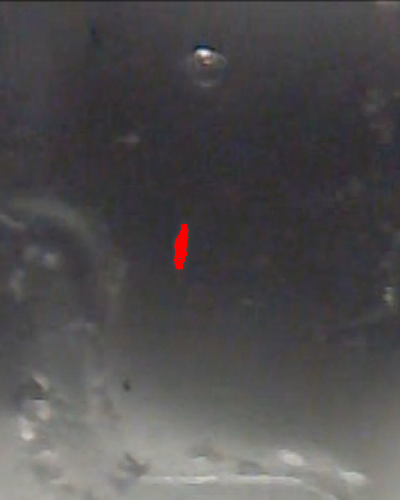}}
\subfigure[]{\includegraphics[width=0.32\textwidth]{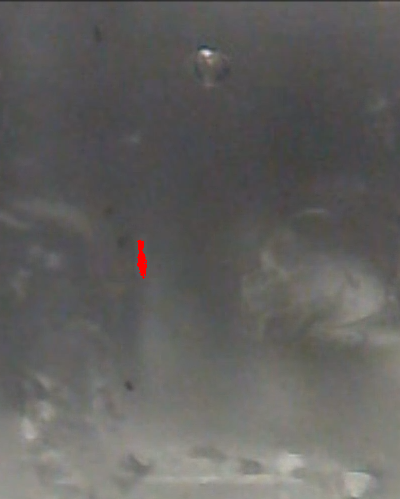}}
\subfigure[]{\includegraphics[width=0.32\textwidth]{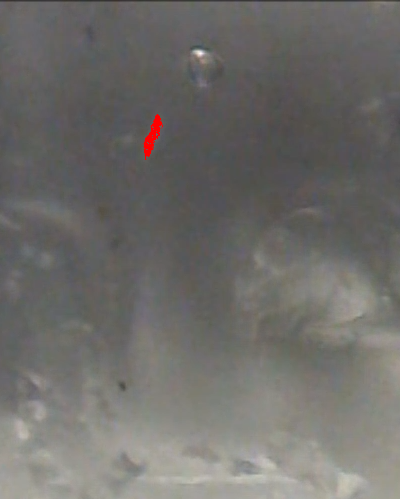}}
\subfigure[]{\includegraphics[width=0.32\textwidth]{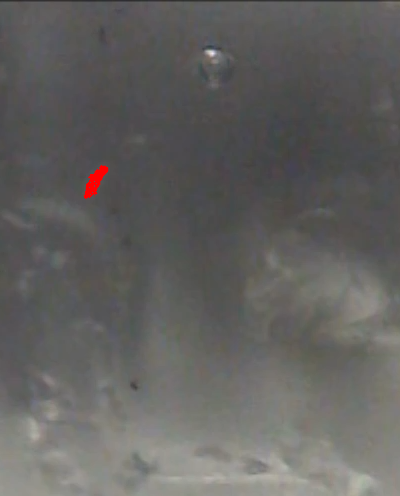}}
\subfigure[]{\includegraphics[width=0.32\textwidth]{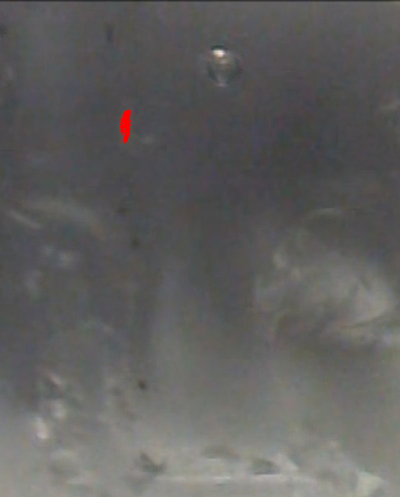}}
\subfigure[]{\includegraphics[width=0.55\textwidth]{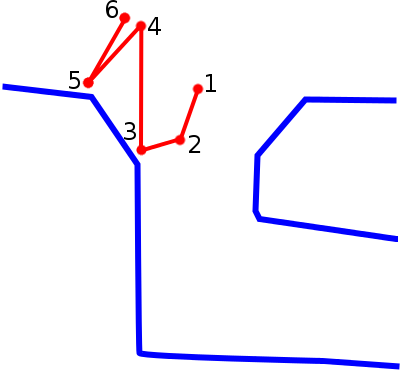}}
\caption{Figure to demonstrate motion of \emph{P. caudatum} in a labyrinth. (a--f) Experimental photographs to show movement of organism (false coloured red) about geometric constraints in a PDMS labyrinth (see text). Images taken at approx. 1~s intervals. Scale bar (in panel $a$) 500~$\mu$m. (g) Schematic to demonstrate organism movement (red, numbered sequentially) about labyrinth boundaries (blue). }
\label{fig-paraMaze}
\end{figure}

\subsection{Results}
The experiment detailed in Fig.\ref{fig-paraMaze} is representative of all experiments conducted. Immediately after inoculation into the maze puzzle, the \emph{P. caudatum} cell spent several seconds rotating on the spot. Following, the organism would migrate in approximately the correct direction, i.e. towards the aperture leading to the maze, before colliding with a wall. Collisions would cause the the organism to migrate in a reverse direction at apparently random angles. This would begin an erratic oscillation in anteroposterior migration which prevented the organism from progressing far into the maze; the furthest a \emph{P. caudatum} cell was observed to have migrated within the 10 minute experiment was approximately 4~mm end-to-end, amounting to two right angle corners successfully traversed.

\subsection{Discussion}
\subsubsection{Spontaneous alternation behaviour is incompatible with constrained-geometry puzzles}
Spontaneous alternation behaviour (SAB) in \emph{Paramecium} spp. is a well documented phenomenon that occurs when the organism collides with a solid object or otherwise meets an unfavourable stimulus. As Witcherman noted in his classic treatise on \emph{Paramecium} \cite{Witcherman1982}, this behaviour is not quite taxis nor quite kinesis as it interferes with true directional movement with a non-specific reaction to move away in any other direction. This response was anticipated, although attempts to engineer the labyrinth's passages as widely as possible had been made as earlier experiments with narrower channels produced much the same effect (data not shown). 

The evolutionary advantages of SAB demonstrate the momentary requirements of motile freshwater microorganisms that sit in the middle of their food chain (i.e. they predate smaller organisms and are prey to larger organisms), as it would appear to favour keeping the organism away from harm at the cost of reducing the efficiency of its search for food. This does, however, result in the organism being virtually incapable of efficiently traversing environments such as the miniature labyrinths used in these experiments, although this of course does not imply that they are unable to address MPPs in unconstrained geometries.

Previous literature has indicated that \emph{P. caudatum} responds to extremely confined environments (microfluidic circuitry or otherwise sealed systems) by exhibiting a somersaulting behaviour which allows it to assume sinusoidal paths by bending over on its self, rather than reverting SAB \cite{Jana2015}. It must be noted, however, that this behaviour is quite different to any previously described mode of \emph{Paramecium} movement and is unlikely to occur in anything but the most confined environments.

\subsubsection{Challenges associated with biocomputing in aqueous environments}
Generating a chemoattractant gradient in an aqueous medium is not a straightforward task. Chemical attraction was chosen as the input over, for example, light or electrical gradients due to the technical limitation of microscopic illumination and the necessity to use electrical fields of a potentially harmful magnitude to be detectable at opposite ends of the labyrinth, respectively. It was reasoned that the organisms were able to sense a chemoattractant given their propensity to always initially migrate in the direction of the stimulus. Nutrient density at the labyrinth start point was low and the organisms only exhibited patterns of movement associated with migration, as opposed to the slower-swimming `feeding behaviours' that can be observed in areas of sufficiently high nutrient substrate concentration \cite{Mayne2017}. Nevertheless, it is apparent that in this instance, any chemoattractive effects that were induced were insufficient to guide the organism through the maze without collisions. Representing multiple vertices as attractant fields in unconstrained geometric spaces would represent a significant technical challenge.

Another issue encountered, which is not adequately represented by standard photography, was the impact that being in a three dimensional environment had on the organisms' pathfinding abilities: \emph{P. caudatum} cells were observed to collide with both the upper and lower levels of the experimental environment (the `roof' and `ceiling', as demarcated by the area filled with fluid) in the majority of experiments, to much the same effect as wall collisions. Even though these labyrinth puzzles are pseudo-two dimensional puzzles, it must be remembered that the search for an efficient path encompasses the need to find the most efficient route in all three directions. Even had the organisms successfully navigated to the labyrinth exit, care would have to be taken in stating that the organism had navigated the `shortest path' through, due to the phenomenon of helical swimming in all \emph{Paramecium} species and z-axis deviations \cite{Jana2012}. This emphasises the need to take into account the third dimension when conducting such experiments.

\subsection{Summary}
Despite being an intensely researched-upon model organism that exhibits many behaviours that can be interpreted as expressions of natural computing (orchestrated manipulation of micro-scale objects \cite{Mayne2017}, basic learning/memory \cite{Hennessey1979}), their mode of motion and hazard avoidance essentially precludes their use in experiments which use geometries constrained to the degrees described. This implies that the only way \emph{P. caudatum} may be coaxed into addressing MPPs is either in architectureless space (which is complex to monitor microscopically) or otherwise in extremely confined environments, which would in turn still be problematic to achieve in practice. 

Although it could perhaps have been predicted that \emph{P. caudatum} cells are a good medium for implementing graph theory biocomputation, it is nevertheless a fact that \emph{Paramecium} species have long-since been used as model aquatic organisms for investigating various taxes, meaning that results gained with these organisms may be partially representative of a large class of organisms. Although there are doubtless other species of ciliate better suited to such applications, this section serves to emphasise that being placed in confined environments is not representative of \emph{P. caudatum's} natural habitat (i.e. large bodies of static or running freshwater) and by extension that live substrates are not always as tolerant to abuse (in designing MPPs or any other form of biocomputer) as archetypal substrates such as slime mould may suggest.

\section{Conclusions}
It is a singular temptation to view emergent biological phenomena in our world and dream of how they may be harnessed, mimicked or emulated for computing applications. Whilst I have endeavoured to prevent the tone of this chapter from being overtly negative or unduly sceptical (biocomputing is a wonderful science whose advancement is of great necessity!), I have attempted to highlight the experimental considerations that make laboratory experiments involving morphological and topological operations with live substrates difficult to implement and even harder to interpret, reproduce and refine. The data presented here suggest that shortest path biocomputation is not something that can easily be achieved and that a certain amount of optimism and open-mindedness is required to interpret such experiments as anything other than rough approximations of MPP solutions in the rare experiments that are ostensibly successful.

In conclusion, the experiments outlined here involve taking organisms out of their natural environment and encouraging them to do distinctly unnatural things. Interpreting the output of a biocomputer almost always requires a certain degree of abstraction; this does not devalue the experiments or make the phenomena under investigation less interesting, but it must be remembered that approximation is a distinctly \emph{unconventional} paradigm in computer science. The best biocomputing experiments will put minimal stress on the cell and observe processes that occur naturally, only intervening subtly in order to tweak a parameter or make a measurement.  

Finally, I will note on behalf of all experimentalists in the field that the divergence between the observed results of biocomputation and our pre-conceived notions of their expected outcomes in comparison with the algorithmic equivalents highlights why it is essential to make physical biocomputing prototypes in unconventional computing research, rather than computer models alone.

\begin{acknowledgement}
The author thanks both reviewers for their invaluable insights and suggestions and Matthew Hynam for providing the laser cut architectural drafts used in slime mould experiments.
\end{acknowledgement}

\section*{Appendix}

\bibliographystyle{spmpsci}
\bibliography{references}

\begin{thebibliography}{10}
\providecommand{\url}[1]{{#1}}
\providecommand{\urlprefix}{URL }
\expandafter\ifx\csname urlstyle\endcsname\relax
  \providecommand{\doi}[1]{DOI~\discretionary{}{}{}#1}\else
  \providecommand{\doi}{DOI~\discretionary{}{}{}\begingroup
  \urlstyle{rm}\Url}\fi

\bibitem{Adamatzky2007}
Adamatzky, A.: {Physarum machines: encapsulating reaction-diffusion to compute
  spanning tree.}
\newblock Naturwissenschaften \textbf{94}(12), 975--80 (2007).
\newblock \doi{10.1007/s00114-007-0276-5}.
\newblock \urlprefix\url{http://www.ncbi.nlm.nih.gov/pubmed/17603779}

\bibitem{Adamatzky2009b}
Adamatzky, A.: {Developing Proximity Graphs By Physarum Polycephalum: Does the
  Plasmodium Follow the Toussaint Hierarchy?}
\newblock Parallel Processing Letters \textbf{19}(1), 105--127 (2009).
\newblock \doi{10.1017/CBO9781107415324.004}

\bibitem{Adamatzky2010}
Adamatzky, A.: {Physarum Machines: Computers from Slime Mould}.
\newblock World Scientific (2010)

\bibitem{Adamatzky2013}
Adamatzky, A.: {Towards slime mould colour sensor: Recognition of colours by
  Physarum polycephalum}.
\newblock Organic Electronics \textbf{14}(12), 3355--3361 (2013).
\newblock \doi{10.1016/j.orgel.2013.10.004}

\bibitem{Adamatzky2016}
Adamatzky, A. (ed.): {Advances in Physarum Machines}.
\newblock Springer International Publishing (2016).
\newblock \doi{10.1007/978-3-319-26662-6}

\bibitem{Adamatzky2014}
Adamatzky, A., Armstrong, R., {De Lacy Costello}, B., Deng, Y., Jones, J.,
  Mayne, R., Schubert, T., Sirakoulis, G., Zhang, X.: {Slime mould analogue
  models of space exploration and planet colonisation}.
\newblock Journal of the British Interplanetary Society \textbf{67}, 290--304
  (2014)

\bibitem{Baumgardner2009}
Baumgardner, J., Acker, K., Adefuye, O., Crowley, S.T., DeLoache, W., Dickson,
  J.O., Heard, L., Martens, A.T., Morton, N., Ritter, M., Shoecraft, A.,
  Treece, J., Unzicker, M., Valencia, A., Waters, M., Campbell, A., Heyer,
  L.J., Poet, J.L., Eckdahl, T.T.: Solving a hamiltonian path problem with a
  bacterial computer.
\newblock Journal of Biological Engineering \textbf{3}(1), 11 (2009).
\newblock \doi{10.1186/1754-1611-3-11}

\bibitem{Dorigo1997}
Dorigo, M., Gambardella, L.M.: Ant colony system: a cooperative learning
  approach to the traveling salesman problem.
\newblock IEEE Transactions on Evolutionary Computation \textbf{1}(1), 53--66
  (1997).
\newblock \doi{10.1109/4235.585892}

\bibitem{Esau2014}
Esau, M., Rozema, M., Zhang, T.H., Zeng, D., Chiu, S., Kwan, R., Moorhouse, C.,
  Murray, C., Tseng, N.t., Ridgway, D., Sauvageau, D., Ellison, M.: {Solving a
  Four-Destination Traveling Salesman Problem Using Escherichia coli Cells As
  Biocomputers}.
\newblock ACS Synthetic Biology \textbf{3}, 972--975 (2014).
\newblock \doi{10.1021/sb5000466}

\bibitem{Harvey2009}
Harvey, C.D., Collman, F., Dombeck, D.A., Tank, D.W.: {Intracellular dynamics
  of hippocampal place cells during virtual navigation.}
\newblock Nature \textbf{461}(7266), 941--6 (2009).
\newblock \doi{10.1038/nature08499}

\bibitem{Hennessey1979}
Hennessey, T., Rucker, W., Mcdiarmid, C.: Classical conditioning in paramecia.
\newblock Animal learning and behaviour \textbf{7}(4), 417--423 (1979)

\bibitem{Jana2015}
Jana, S., Eddins, A., Spoon, C., Jung, S.: {Somersault of Paramecium in
  extremely confined environments}.
\newblock Scientific Reports \textbf{5}, 13,148 (2015).
\newblock \doi{10.1038/srep13148}

\bibitem{Jana2012}
Jana, S., Um, S.H., Jung, S.: {Paramecium swimming in capillary tube}.
\newblock Physics of Fluids \textbf{24}(4) (2012).
\newblock \doi{10.1063/1.4704792}

\bibitem{Jones2015}
Jones, J.: Mechanisms inducing parallel computation in a model of physarum
  polycephalum transport networks.
\newblock Parallel processing Letters \textbf{25}, 1540,004 (2015).
\newblock \doi{10.1142/S0129626415400046}

\bibitem{Jones2015b}
Jones, J., Mayne, R., Adamatzky, A.: {Representation of shape mediated by
  environmental stimuli in Physarum polycephalum and a multi-agent model}.
\newblock International Journal of Parallel, Emergent and Distributed Systems
  (2), 166--184 (2015).
\newblock \doi{10.1080/17445760.2015.1044005}

\bibitem{Mayne2016advances}
Mayne, R.: {Biology of the Physarum polycephalum plasmodium: preliminaries for
  unconventional computing}.
\newblock In: A.~Adamatzky (ed.) Advances in Physarum Machines, vol.~21,
  chap.~1, pp. 3--22. Springer (2016).
\newblock \urlprefix\url{http://link.springer.com/10.1007/978-3-319-26662-6}

\bibitem{Mayne2016}
Mayne, R.: {Orchestrated Biocomputation: Unravelling the Mystery of Slime Mould
  "Intelligence".}
\newblock Luniver Press, Bristol, UK (2016)

\bibitem{Mayne2015}
Mayne, R., Adamatzky, A.: {Slime mould foraging behaviour as optically coupled
  logical operations}.
\newblock International Journal of General Systems \textbf{44}(3), 305--313
  (2015).
\newblock \doi{10.1080/03081079.2014.997528}

\bibitem{Mayne2017}
Mayne, R., Whiting, J.G., Wheway, G., Melhuish, C., Adamatzky, A.: {Particle
  Sorting by Paramecium Cilia Arrays}.
\newblock Biosystems \textbf{156-157}, 46--52 (2017).
\newblock \doi{10.1016/j.biosystems.2017.04.001}

\bibitem{Nakagaki2000}
Nakagaki, T., Yamada, H., Toth, A.: Intelligence: Maze-solving by an amoeboid
  organism.
\newblock Nature \textbf{407}, 470 (2000).
\newblock \doi{10.1038/35035159}

\bibitem{Pratt2002}
Pratt, S.C., Mallon, E.B., Sumpter, D.J.T., Franks, N.R.: {Quorum sensing,
  recruitment, and collective decision-making during colony emigration by the
  ant Leptothorax albipennis}.
\newblock Behavioral Ecology and Sociobiology \textbf{52}(2), 117--127 (2002).
\newblock \doi{10.1007/s00265-002-0487-x}

\bibitem{Ramsch2012}
Ramsch, K., Reid, C., Beekman, M., Middendorf, M.: A mathematical model of
  foraging in a dynamic environment by trail-laying argentine ants.
\newblock Journal of Theoretical Biology \textbf{306}, 32--45 (2012)

\bibitem{Reid2012}
Reid, C.R., Latty, T., Dussutour, A., Beekman, M.: {Slime mold uses an
  externalized spatial "memory" to navigate in complex environments.}
\newblock Proceedings of the National Academy of Sciences of the United States
  of America \textbf{109}(43), 17,490--4 (2012).
\newblock \doi{10.1073/pnas.1215037109}

\bibitem{Reid2011}
Reid, C.R., Sumpter, D.J.T., Beekman, M.: Optimisation in a natural system:
  Argentine ants solve the towers of hanoi.
\newblock Journal of Experimental Biology \textbf{214}(1), 50--58 (2011).
\newblock \doi{10.1242/jeb.048173}

\bibitem{Riedel-Kruse2011}
Riedel-Kruse, I.H., Chung, A.M., Dura, B., Hamilton, A.L., Lee, B.C.: {Design,
  engineering and utility of biotic games}.
\newblock Lab Chip \textbf{11}(1), 14--22 (2011).
\newblock \doi{10.1039/C0LC00399A}

\bibitem{Saiz2006}
Saiz, L., Vilar, J.M.G.: {Stochastic dynamics of macromolecular-assembly
  networks}.
\newblock Molecular Systems Biology pp. 1--11 (2006).
\newblock \doi{10.1038/msb4100061}

\bibitem{Scherber2012}
Scherber, C., Aranyosi, A.J., Kulemann, B., Thayer, S.P., Toner, M.,
  Iliopoulos, O., Irimia, D.: {Epithelial cell guidance by self-generated EGF
  gradients}.
\newblock Integrative Biology \textbf{4}(3), 259 (2012).
\newblock \doi{10.1039/c2ib00106c}

\bibitem{Stepney2008}
Stepney, S.: {The neglected pillar of material computation}.
\newblock Physica D: Nonlinear Phenomena \textbf{237}, 1157--1164 (2008).
\newblock \doi{10.1016/j.physd.2008.01.028}

\bibitem{Tero2010}
Tero, A., Takagi, S., Saigusa, T., Ito, K., Bebber, D.P., Fricker, M.D.,
  Yumiki, K., Kobayashi, R., Nakagaki, T.: {Rules for biologically inspired
  adaptive network design.}
\newblock Science \textbf{327}(5964), 439--42 (2010).
\newblock \doi{10.1126/science.1177894}.
\newblock \urlprefix\url{http://www.ncbi.nlm.nih.gov/pubmed/20093467}

\bibitem{openSCAD}
Thingiverse: openscad maze generator.
\newblock Available at \url{https://www.thingiverse.com/thing:24604} ([Accessed
  01 May 2017).
\newblock Produced by user 'dnewman'

\bibitem{VanHouten1978}
Van~Houten, J.: Two mechanisms of chemotaxis inparamecium.
\newblock Journal of comparative physiology \textbf{127}(2), 167--174 (1978).
\newblock \doi{10.1007/BF01352301}

\bibitem{VanHouten1975}
Van~Houten, J., Hansma, H., Kung, C.: Two quantitative assays for chemotaxis
  inparamecium.
\newblock Journal of comparative physiology \textbf{104}(2), 211--223 (1975).
\newblock \doi{10.1007/BF01379461}

\bibitem{Vela-Perez2013}
Vela-P\'{e}rez, M., Fontelos, M., Vel\'{a}squez, J.: Ant foraging and geodesic
  paths in labyrinths: Analytical and computational results.
\newblock Journal of Theoretical Biology \textbf{320}, 100--112 (2013)

\bibitem{Vittori2006}
Vittori, K., Talbot, G., Gautrais, J., Fourcassie, V., Araujo, A., Theraulaz,
  G.: Path efficiency of ant foraging trails in an artificial network.
\newblock Journal of Theoreical Biology \textbf{239}, 507--515 (2006)

\bibitem{Witcherman1982}
Witcherman, R.: The biology of {P}aramecium, second edn., chap. 5: Movement,
  behaviour and motor response, pp. 211--238.
\newblock Plenum (1982)

\bibitem{Zhu2013}
Zhu, L., Aono, M., Kim, S.J., Hara, M.: {Amoeba-based computing for traveling
  salesman problem: Long-term correlations between spatially separated
  individual cells of Physarum polycephalum}.
\newblock BioSystems \textbf{112}(1), 1--10 (2013).
\newblock \doi{10.1016/j.biosystems.2013.01.008}

\end{thebibliography}

\end{document}